\documentclass[english]{llncs}
\usepackage[T1]{fontenc}
\usepackage[latin9]{inputenc}
\usepackage{amssymb}

\makeatletter
\usepackage{dsfont}

\makeatother

\usepackage{babel}
\begin{document}

\title{On a Model of Quantum Mechanics and the Mind}

\author{J. Acacio de Barros}

\institute{Liberal Studies Program, BH238\\
San Francisco State University\\
1600 Holloway Ave.\\
San Francisco, CA 94132}
\maketitle
\begin{abstract}
In this paper I discuss Stapp's interesting proposal of using the
Quantum Zeno Effect to account for the mind/matter interaction \cite{stapp_mind_2014}.
In particular, I discuss some of the motivations for it, and then
argue that, in his current version, his model is circular (a solution
to this, proposed by Kathryn Laskey, is presented), insofar as the
mind/matter problem is concerned. I also present an alternative approach
to some of the appealing aspects of using quantum mechanics to think
about consciousness. 
\end{abstract}

\section{Introduction}

This paper is an extension of the main points I started to discuss
during the Foundations of Mind Conference, organized by Sean O'Nuallain
and held in Berkeley in March 2014. I was fortunate to be invited
by Sean to chair a section where Henry Stapp would talk about his
theory of quantum mechanics and the mind. But Sean also gave me the
almost impossible task of criticizing Stapp's views. Of course, as
soon as I realized that one can criticize without actually giving
a plausible alternative, the task became less daunting. So, the goal
of this paper is to put forth a couple of criticisms of the quantum
mind theory and, perhaps, suggest a possible alternative of what theory
could replace at least certain aspects of it. 

Before going on, let me start with a general comment. I do not wish
to claim here that Stapp's theory of quantum mind is without merits.
I find it a fascinating way to think about the mind and its connection
to quantum mechanics, a subject that has fascinated me ever since
I was a graduate student and got in touch with many texts on the subject,
including von Neunann's seminal book \cite{von_neumann_mathematical_1996}.
My goal here is to point out some difficulties that I see with the
theory, both from a conceptual as well as a technical point of view,
and to question it as a way to help with the problem of consciousness.
This will be done, hopefully, without resorting to too many technical
arguments, and some liberty will be taken about some details. I will,
however, assume that the reader is familiar with quantum mechanics
and its formalism and notation at the level of standard textbooks,
such as \cite{cohen-tannoudji_quantum_1977}. 

I organize this paper in the following way. The first part of my argument
is presented in section \ref{sec:von-Neumann's-Approach}, I discuss
the von Neumann approach to quantum mechanics, the backbone of Stapp's
theory. In it, I try to argue that von Neumann's views, albeit \emph{consistent}
with empirical evidence, loose their motivation when viewed from a
modern perspective. Motivation, of course, is in the eyes of the beholder,
and in this section I cannot hope but only convince those who were
already skeptical of the quantum-mind theory. However, revisiting
von Neumann's theory is useful for my more focused criticisms, spelled
out in the next section. In section \ref{sec:Stapp's-quantum-mind}
I quickly introduce some main aspects of Stapp's theory (detailed
in \cite{stapp_mind_2014} on this proceedings), as I (hopefully)
understand it, and I present two arguments against it. One argument
is more specific to the particular application of QZE shown in \cite{stapp_mind_2014}.
The other argument is broader, and questions the proposed use of QZE
to solve the mind/matter problem. In particular, we show that Stapp's
use of QZE leads to a circularity in his argument, and therefore is
not really a solution to the mind/matter problem, though it can be
modified to account for this. In section \ref{sec:Possible-Alternatives}
I end the paper on a more positive note, giving some possible alternative
approaches that are somewhat in the spirit of what Stapp is trying
to achieve with his theory of the quantum mind.

\section{von Neumann's Approach\label{sec:von-Neumann's-Approach}}

Let me start with the overall ``historic'' argument about the connection
between the mind and quantum mechanics, first put forth by von Neumann
himself in his discussions of the theory of measurement \cite{von_neumann_mathematical_1996}.
According to the Copenhagen interpretation, one of the difficulties
of quantum mechanics, first pointed out clearly by Bohr \cite{bohr_quantum_1928},
is the dual nature of the evolution of a physical system. On the one
hand quantum mechanics is deterministic: given the state $|\psi_{0}\rangle$
of a system at time $t_{0}$, its time evolution is given by 
\begin{equation}
|\psi\rangle=e^{-\frac{i}{\hbar}\hat{H}t}|\psi_{0}\rangle,\label{eq:evolution}
\end{equation}
where $\hat{H}$ is the Hamiltonian operator. Equation (\ref{eq:evolution})
uniquely determines the system $|\psi\rangle$ at time $t\geq t_{0}$.
On the other hand, quantum mechanics, through the measurement process,
is probabilistic: all we can talk about from the $|\psi\rangle$ are
the probabilities $P\left(o_{i}\right)=\left|\langle o_{i}|\psi\rangle\right|^{2}$
of possible outcomes $o_{i}$ of an experiment $\mathcal{O}$ represented
by the observable $\hat{O}$, such that $\hat{O}|o_{i}\rangle=o_{i}|o_{i}\rangle$.
So, it seems that there are two different types of incompatible evolution
in quantum mechanics, and one of them is associated to a very special
type of interaction: a measurement.

The question posed by the founders of quantum mechanics was the following:
what makes measurements different? To Bohr, a measurement was an interaction
with a classical system. This is all good when we are talking about,
say, an electron (clearly quantum) and dark spots on a photographic
paper (clearly classical). But the problem becomes trickier when the
measurement device itself is small: for instance, when we think of
atoms as photodetectors, or when we deal with mesoscopic systems.
Some measurement devices are clearly classical, whereas others are
not. So, where is this boundary between classical and quantum physics?
This border is what was known as the Heisenberg cut, the point where
anything over it behaves classically. 

Von Neumann considered this idea of a dual dynamics, one for classical
measurement apparatus and another for quantum systems, unsatisfactory
\cite{von_neumann_mathematical_1996}. To overcome this, he treated
both the observed system \emph{and} the measurement apparatus as quantum
systems. A measurement consisted thus of something with the following
characteristic. Let $\hat{O}$ be the observable represented by the
(quantum-mechanically described) measuring apparatus $\mathcal{O}$,
such that if a system is initially in the state $|o_{i}\rangle$ and
the apparatus in the state $|0\rangle$ (i.e., the state is represented
by the value of the pointer on the measuring device measuring nothing,
the reset position), then the interaction leads the the following
evolution:
\[
|o_{i}\rangle\otimes|0\rangle\rightarrow|o_{i}\rangle\otimes|i\rangle,
\]
where $|i\rangle$ means the apparatus is pointing to the value corresponding
to the outcome $o_{i}$. This is all fine with states for the system
that are eigenstates of the measurement apparatus. However, if the
system is in a superposition of the type 
\[
|\psi\rangle=c_{1}|o_{1}\rangle+c_{2}|o_{2}\rangle,
\]
$\left|c_{1}\right|^{2}+\left|c_{2}\right|^{2}=1$, the interaction
with the measurement apparatus leads to 
\[
|\psi\rangle\otimes|0\rangle\rightarrow c_{1}|o_{1}\rangle\otimes|1\rangle+c_{2}|o_{2}\rangle\otimes|2\rangle.
\]
Notice that this last equation is not what happens with a measurement
process, where we either get one outcome or the other. Instead, because
of the linearity of quantum evolution, superpositions of the initial
state lead to, after an interaction with the measurement device, a
quantum superposition of the whole system. In other words, there is
no collapse of the wave function, and therefore, no ``measurement'',
in the sense of Bohr, happened. From the linearity of quantum mechanics,
von Neumann observed that the quantum superposition of system and
apparatus could be extended all the way to larger and larger systems.
However, there was one point where unambiguously there was no superposition:
the mind of an observer. This is the case because we \emph{never}
see the superposition of, say, Schrödinger's cat dead and alive. So,
von Neumann posited that the actual irreversible aspect of the quantum
measurement, the collapse of the wave function, happens at the mind,
thus avoiding the issue of determining Heisenberg's cut. 

Thus, von Neumann's idea that the Heisenberg cut, i.e. where does
quantum interference is lost and classical outcomes exist, is done
in the interaction between the mind and the systems correlated to
the original quantum system being measured. For example, an atom is
detected by a detector, which is then observed by the researcher's
eyes, which send signals to the brain. At each step, atom, detector,
and so on, we have a chain of entangled quantum systems. It is not
until this chain interacts with the mind (somewhere after the eyes)
that a collapse of the wave function actually happens. 

The main motivation for von Neumann to go ``all the way up'' to
the mind was the lack of clear boundary between the classical and
the quantum regimes. However, as we know from modern environmental
decoherence theory%
\footnote{Some might point that von Neumann already accounted for decoherence
in his book, as he talks about the possibility for off-diagonal elements
of the partial trace of the density matrix to go to zero because of
a measurement process. We clarify that what we are talking about here
is the dynamical theory of decoherence that, for instance, makes explicit
claims about how the off-diagonal terms of the density matrix go to
zero, a result that has practical implications for the construction
of quantum computers. This theory's predictions have been observed,
for example, in mesoscopic systems \cite{brune_observing_1996}. %
}, such boundary is not as muddled as initially thought \cite{omnes_interpretation_1994}.
Furthermore, quantum superposition, the cornerstone of quantum effects,
is not the norm for larger systems, and interference effects associated
to such superpositions decay extremely rapidly even for mesoscopic
systems, in accordance with the theory \cite{brune_observing_1996}. 

Now, what does environmental decoherence has to tell us about von
Neumann's quantum mind? First of all, it does not solve the measurement
problem, the main reason that led von Neumann to his interpretation,
even if we accept Bell's concept of solving For All Practical Purposes
(FAPP). Because environmental decoherence still relies on quantum
evolution, it still carries all the way up the same superpositions
that have troubled physicists for a century. 

Second of all, though decoherence clarifies the Heisenberg cut, it
is by no means a disproof of von Neumann's theory. Because the underlying
dynamics is still linear, one can argue that quantum superpositions
still exist, and that the ``collapse'' only happens in the mind%
\footnote{This is possibly the main reason why so many of the proponents of
decoherence as a way to clarify the measurement problem are sympathetic
to the many worlds/minds interpretation of quantum mechanics. %
}. If not anything else, decoherence probably makes it impossible to
falsify von Neumann's ideas, as there is no way quantum coherence
can be held up to the, say, brain level \cite{tegmark_importance_2000},
in an observable way. 

But decoherence does tell us that we can \emph{talk} about the Moon,
for all practical purposes, even if we have not observed it, and explains
why most macroscopic things behave in a seemingly classical way. So,
in a certain sense, it makes von Neumann's theory more far-fetched.
Notice that von Neumann proposed a solution to the measurement problem
by denying a dual dynamics of quantum and classical evolution for
different systems, but instead by creating a new system, not subject
to the laws of quantum mechanics, that had its own dynamics as well:
the mind. So, it simply replaces a mystery by another mystery, without
adding any explanatory power.

More importantly, decoherence removes the main motivation for von
Neumann's collapse postulate. If we recall, this postulate was introduced
(by Heisenberg) to deal with the fact that we do not see macroscopic
quantum superpositions. But let us say that we do indeed have, as
von Neumann says, a macroscopic quantum superposition getting all
the way to the mind. Everett's many worlds/minds interpretation simply
says that our mind makes a selection, among the many possible, but
we still have a superposition \cite{everett_relative_1957}. What
selection does the mind make? Not any of the infinite amount of possible
experimental outcomes, but instead such selections that are consistent
with the preferred pointer basis of the involved quantum observables,
as given by the decoherence of the quantum states. So, no need for
any special dynamics of the mind outside of the quantum (or physical)
realm.

\section{Stapp's Quantum Mind\label{sec:Stapp's-quantum-mind}}

I now turn to Stapp's approach \cite{stapp_mind_2009}, more specifically
his arguments laid out in \cite{stapp_mind_2014}. Let us assume that
von Neumann's theory is correct, and that there is a different entity,
the mind, that does not satisfy the laws of quantum mechanics, and
is responsible for the classical character of the measurement. Stapp
poses the interesting question of whether such entity could affect
the physical world, a huge problem for the proponents of the mind/brain
duality. 

Stapp suggests the Quantum Zeno Effect (QZE) \cite{misra_zenos_2008}
as a possible mechanism for the mind to affect matter. In the original
QZE, it was shown that if we continuously observed an unstable particle,
this particle would not decay. However, we can modify this argument,
and show that by continuous observations (or many observations close
to each other) we can make a particle change a quantum state. To use
Stapp's example in \cite{stapp_mind_2014}, let us start with a coherent
state with amplitude $\alpha$, represented by the ket $|\alpha\rangle$.
Now, let us perform a simple yes/no experiment, where the question
being asked is whether the system is a coherent state with amplitude
$\alpha+\Delta$, where $\Delta$ is small compared to $\alpha$.
Because coherent systems are not orthonormal, the probability for
the state $|\alpha\rangle$ to be in $|\alpha+\Delta\rangle$ is nonzero,
and given by 
\begin{equation}
P\left(|\alpha+\Delta\rangle||\alpha\rangle\right)=\left|\langle\alpha|\alpha+\Delta\rangle\right|^{2}\approx1-\Delta^{2}\label{eq:prob-alphaplusdelta}
\end{equation}
for $\Delta^{2}$ sufficiently small. Furthermore, if we make $N$
several successive measures, each time asking the question with a
larger amplitude (by a factor $\Delta$), the probability of observing
$\alpha+N\Delta$ is given by 
\[
P\left(\alpha+N\Delta\right)\approx\left(1-\Delta^{2}\right)^{N}\approx1-N\Delta^{2},
\]
and \emph{not }$\left|\langle\alpha|\alpha+N\Delta\rangle\right|^{2}$.
Thus, making $\Delta$ very small, i.e. by almost continuously observing
a quantum harmonic oscillator in the semi-classical coherent state,
it is possible to increase its amplitude of oscillation. 

In \cite{stapp_mind_2014}, Stapp applies this ideas to motor cortex
measurements performed by Rubino, Robbins, and Hatsopoulos \cite{rubino_propagating_2006}.
His idea is that in the same way that the mind causes the collapse
of the wave function, the effect of the mind ``observing'' a system
can make it change its state from $|\alpha\rangle$ to $|\alpha+N\Delta\rangle$.
I see two problems with this model, one conceptual and one specific
to the application. 

The specific problem is more technical in nature, and perhaps not
as important as the second, so let me talk about it first. Stapp starts
with the magnetic field in what he claims is a computational unit:
the minicollumn. Because the magnetic field is very weak, of the order
of pT, it follows that if we model its oscillations with a quantum
harmonic oscillator in a coherent state (as the computations above),
it is justifiable to use quantum mechanics, as $\alpha$ is only at
the order of $10^{1}$. However, why talk about the magnetic field?
Given the low frequencies, the magnetic and electric fields are uncoupled,
and the regime is essentially a quasi-static one \cite{nunez_electric_2006}.
Furthermore, as \cite{rubino_propagating_2006} show, the electric
field involved in the process is at the order of $10$ $\mu\mbox{V}$,
and its energy is about $10$ orders of magnitude that of the magnetic
field. More importantly, because it is a much stronger field, it not
only is more relevant to understand the processes, but it is also
describable, to a very good approximation, by classical equations,
as one should expect. So, to make his model stick, Stapp would have
to clearly justify the picking of magnetic fields over electric fields,
and more importantly, how such fields, when increased even tenfold,
would affect the dynamics of the brain in a significant way (given
that it carries energy that is about $10^{9}$ times less than the
electric field). 

We now turn to the main problem with Stapp's approach. As we mentioned
above, Stapp starts with a weak magnetic field, modeled by a coherent
state, which is subjected to successive measurements. The mind is
seen affect the field itself through the QZE, a (perhaps measurable)
mechanism for the mind/brain interaction. In other words, if the mind
chooses to measure $|\alpha+\Delta\rangle$, and then $|\alpha+2\Delta\rangle$,
then $|\alpha+3\Delta\rangle$, and so on, then it can make the amplitude
of the field increase, within a reasonable amount of time, from $\alpha$
to $\alpha+N\Delta$. 

But what is a measurement of $|\alpha+\Delta\rangle$ or $|\alpha+2\Delta\rangle$?
In von Neumann's formulation, this process is done by the presence
of a physical system, i.e., by some hardware (responsible, as shown
by decoherence, for the choice of a pointer basis). For example, to
measure the spin of an electron, we have to produce a Stern-Gerlach
experimental setup before the atoms get to the photographic place,
which then sends photons to the eyes of the observer, that activate
neurons in the brain, and somewhere or somehow finally gets to the
mind. But the Stern-Gerlach setup needs to be there; it cannot be
produced by the mind. The same is true for the QZE. A \emph{physical
}measurement has to be made to affect the system. 

What the mind does is only, according to von Neumann, collapse the
wave function. It does not make a measurement, as the mind does not
have a preferred pointer basis, which is itself provided by the theory
of decoherence through super selection rules \cite{zurek_environment-induced_1982}.
In Stapp's model, the mind affects matter by measuring first, say,
$|\alpha+\Delta\rangle$, and then $|\alpha+2\Delta\rangle$, and
so on. But the choice of measuring $|\alpha+2\Delta\rangle$ instead
of $|\alpha+\Delta\rangle$ involves the presence of a physical apparatus
that measures $|\alpha+2\Delta\rangle$, physically different from
the apparatus that measures $|\alpha+\Delta\rangle$. Such mind may
be ``observing'' this apparatus and making it collapse into one
of its pointer basis, with values ``yes'' or ``no,'' but it cannot
make the measurement itself. Thus, we reach the conclusion that, for
the measurement to be performed, there must be a way for the mind
to affect matter by selecting a specific apparatus and its corresponding
pointer basis (instead of another), and we get into a circular argument:
to solve the problem of how the mind affects matter, we need to postulate
that the mind affects matter. 

The solution to the circularity, which is not explicitly stated in
Stapp's paper \cite{stapp_mind_2014}, was suggested by Kathryn Laskey
(private communication). Imagine that the coherent state $|\alpha\rangle$
is constantly being measured by an apparatus set up by the brain,
in such a way that the apparatus evolves with a cyclic dynamics, such
that at time $t$ the apparatus superposes the system into two possible
states given by the basis $|\alpha+\gamma\sin\left(\omega t\right)\rangle\langle\alpha+\gamma\sin\left(\omega t\right)|$
or $\hat{\mathds{1}}-|\alpha+\gamma\sin\left(\omega t\right)\rangle\langle\alpha+\gamma\sin\left(\omega t\right)|$.
The \emph{mind, }in this case, can choose to observe or not the system,
and can also choose when to observe the system. It the mind does not
observe, no changes to the coherent state happen. If, on the other
hand, the mind continuously observes the system between a time $t'$,
$2\pi n\leq t'\leq2\pi n+\frac{\pi}{2}$, an irreversible change happens,
with complete loss of coherence, and the system ends in the state
$|\alpha+\gamma\rangle$ (notice that a continuous observation leads
to a probability one for the QZE). This solution to the circularity
argument presents an interesting case, in which the quantum interaction
of the mind with matter would be the consequence of a particular type
of brain evolution (of course, not necessarily the $\sin$ oscillation,
but any cyclic system would do), which could in principle be observable.

\section{Possible Alternatives\label{sec:Possible-Alternatives}}

I end this paper with possible alternatives on how to approach the
issue of consciousness (the easy problem, of course). As Stapp mentioned
in a private conversation, one of his motivations for the quantum
theory of mind was that it provides an alternative to human decision
making being either random (say, as modeled by the classical SR behaviorist
theory) or deterministic. He claimed that quantum mechanics, through
von Neumann's interpretation, provided an third way: how to think
about the decision maker as a free agent (thus non-deterministic)
while at the same time not simply coming to decisions by a random
process. 

Though either alternative seems almost like a (free?) personal choice
on how one wants to think about the relationship between consciousness
and decision making, I want to provide a fourth way. In between deterministic
and stochastic processes, may be stochastically incomplete processes.
We know that decision making is highly contextual, in the sense that
the probabilistic processes involved in many human decisions cannot
be appropriately modeled by classical probability theory \cite{kahneman_perspective_2003}.
In fact, this stochastic incompleteness is closely related to the
quantum mechanical one, in the sense that certain decision-making
processes can be better modeled by an algebraic structure inspired
by quantum mechanics (see \cite{suppes_quantum_2007,khrennikov_quantum-like_2009,de_barros_quantum_2009,khrennikov_ubiquitous_2010,haven_quantum_2013,busemeyer_quantum_2012}
and references therein). The fact that the mathematical formalism
of quantum mechanics leads to better descriptions of social phenomena
led to the term \emph{quantum-like} and to new areas of research,
such as quantum cognition and quantum finances. 

However, as pointed out by many authors \cite{suppes_quantum_2007,khrennikov_quantum-like_2009,de_barros_quantum_2009,khrennikov_ubiquitous_2010,haven_quantum_2013,busemeyer_quantum_2012},
this quantum-like behavior does not mean that such systems are quantum
mechanical. What is meant here is simply that the underlying dynamics
can be described as a classical one, and yet result in quantum-like
effects. In fact, quantum like effects in the brain can be obtained
by simple contextual interference \cite{de_barros_quantum-like_2012,de_barros_beyond_2013},
based on models of neural oscillators that reproduces standard SR
theory in certain cases \cite{vassilieva_learning_2011,suppes_phase-oscillator_2012,de_barros_response_2014}.
This should not be surprising, as realistic models can reproduce the
same outcomes of quantum mechanics, as long as no spacelike events
are involved or if detectors are not 100\% efficient (see \cite{suppes_random-walk_1994,suppes_diffraction_1994,suppes_proposed_1996,suppes_violation_1996,suppes_particle_1996}
for one such model). 

What is at the core of such quantum-like effects is contextuality.
One should not expect social systems to be non-local, as the EPR example,
but one should expect them to be contextual. However, if contextuality
is the name of the game, then perhaps the quantum mechanical apparatus
carries too much baggage with it. For example, one can have stochastically
incomplete systems and yet have no quantum description for them \cite{de_barros_joint_2012}.
Furthermore, such systems can be modeled by neural oscillators, thus
leaving open the idea that using a quantum mechanical description
is too constraining. But, more importantly, with alternative descriptions,
it is possible to show that further principles can be added, providing
a possible constraint in decision making that is neither quantum nor
random \cite{de_barros_decision_2013}. This opens up the exciting
perspective of having an approach to decision making that satisfies
Stapp's criteria, in a certain sense, and is at the same time testable
at the behavioral level. 

Let me end with a final general comment about using classical approaches
to understand something that is, in essence, a quantum phenomena,
as quantum mechanics is the ultimate theory of Nature. Physics is
not only about constructing theories, but also models. For example,
though classical mechanics is essentially wrong, one would not use
QM in a model for spaceship trajectories. A quantum model would not
only be impractical, but would also not add anything to the \textquotedblleft story,\textquotedblright{}
to our understanding of the issue. In fact, even the simplest attempts
to prove the stability of matter from QM have failed miserably at
macroscopic levels (see reference \cite{lieb_stability_1991} for
a review). Models tell us a story of causal (including probabilistically
causal) connections. This is what we call understanding in physics
(e.g., we understand planetary trajectories because we can tell them
from Newton\textquoteright s gravity). More importantly, classical
mechanics is a good approximation for QM for most macroscopic objects
(including not so macroscopic ones such as neurons). In fact, in many
simulations with molecules, Newtonian mechanics works pretty good.
What is fundamental then? QFT? Should we try to abandon even the view
that there are particles in the brain, and try to model it with fields?
Strings? From a pragmatic point of view, we don\textquoteright t use
theories because they are more fundamental. We use approximations
that allow us to say something about the system in a coherent way,
and try to justify such approximations based on reasonable assumptions
and empirical data. I hope the theory put forth in the section, though
classical, may be of help to elucidate certain aspects of decision
making that have eluded explanation.

\paragraph{Acknowledgments. }

I would like to thank Sean O'Nuallain and Henry Stapp for the stimulating
discussions about the connection between quantum mechanics and the
mind, and also Carlos Montemayor and Gary Oas for helpful comments
and suggestions about this manuscript. 

\bibliographystyle{plain}
\bibliography{Quantum}

\end{document}